\documentclass[pdflatex,sn-mathphys-ay]{sn-jnl}

\usepackage{graphicx}%
\usepackage{subfig}
\usepackage{multirow}%
\usepackage{amsmath,amssymb,amsfonts}%
\usepackage{amsthm}%
\usepackage{mathrsfs}%
\usepackage[title]{appendix}%
\usepackage{xcolor}%
\usepackage{textcomp}%
\usepackage{manyfoot}%
\usepackage{booktabs}%
\usepackage{algorithm}%
\usepackage{algorithmicx}%
\usepackage{algpseudocode}%
\usepackage{listings}%

\theoremstyle{thmstyleone}%
\theoremstyle{thmstyletwo}%

\theoremstyle{thmstylethree}%

\begin{document}

\title[Article Title]{Model-Based Cortisol Replacement Strategies for Patients With Adrenal Insufficiency  }


\author*[1]{\fnm{Nayana} \sur{Saha}}\email{nayana.officialid@gmail.com}

\author[1]{\fnm{Renuka} \sur{Joshi}}\email{joshirenuka210@gmail.com}

\author[1]{\fnm{Vittal} \sur{Srinivasan}}\email{srini133@purdue.edu}

\author[1]{\fnm{Stanislaw H.} \sur{\.{Z}ak}}\email{zak@purdue.edu}

\author[2]{\fnm{Cary N.} \sur{Mariash}}\email{cmariash@iu.edu}

\affil*[1]{\orgdiv{Elmore Family School of Electrical and Computer Engineering}, \orgname{Purdue University}, \orgaddress{\street{610 Purdue Mall}, \city{West Lafayette}, \postcode{47906}, \state{Indiana}, \country{United States}}}

\affil[2]{\orgdiv{Division of Endocrinology}, \orgname{Indiana University School of Medicine}, \orgaddress{\street{340
West 10th Street}, \city{Indianapolis}, \postcode{46202}, \state{Indiana}, \country{United States}}}



\abstract{Cortisol is an essential hormone secreted by the adrenal glands. Abnormally low level of cortisol is referred to as adrenal insufficiency. This paper focuses on devising treatment strategies for patients suffering from adrenal insufficiency using a model-based approach.   Two conditions of adrenal insufficiency are analysed. They are hypothalamic-pituitary (HP) and adrenal gland disorders. The hypothalamus, pituitary, and adrenal glands constitute the hypothalamus-pituitary-adrenal (HPA) axis, which regulates stress and maintains physiological homeostasis through cortisol secretion. A comparative analysis of the existing mathematical models of the HPA axis is conducted.  Based on this analysis, a novel model of the HPA axis is proposed that incorporates endogenous circadian rhythm. This proposed model is used to design effective cortisol replacement strategies. In addition, hydrocortisone replacement therapies are proposed and validated with clinical data from the literature. The results presented in this paper are applicable to patient-specific treatment strategies for individuals with adrenal insufficiency.}

\keywords{Endocrine system, Hypothalamus-pituitary-adrenal (HPA) axis, mathematical modeling, circadian rhythm, primary adrenal insufficiency, secondary adrenal insufficiency.}

\maketitle
\begin{figure}[!t]
\centerline{\includegraphics[width=0.75\textwidth]{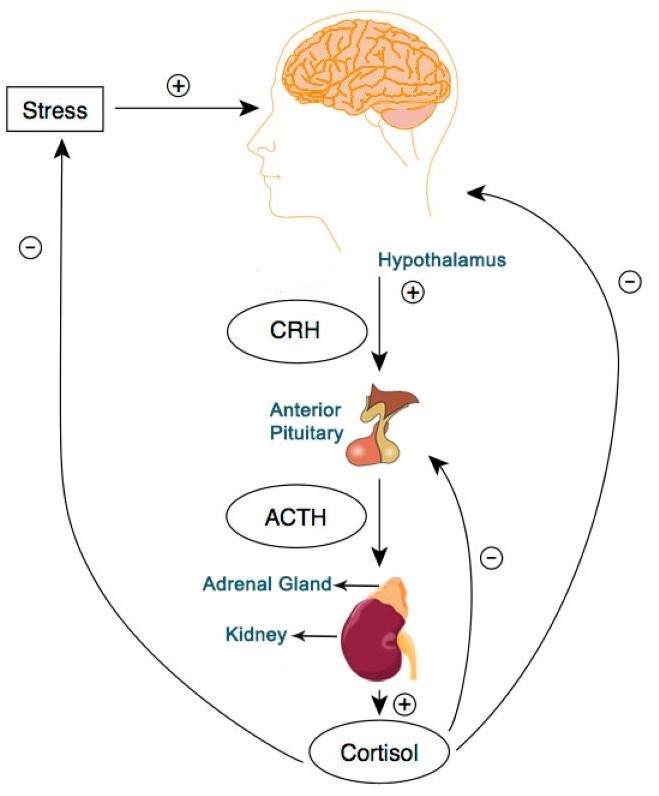}}
\caption{The hypothalamus-pituitary-adrenal (HPA) axis. (Source: Zellx)}~\citep{ZellxHPA2023}.\label{fig:HPA-axis}
\end{figure}

\section{Introduction}\label{sec1}

Adrenal Insufficiency (AI) is a condition in which adrenal glands do not produce sufficient cortisol. This affects the body's ability to regulate and manage stress with altered metabolism and immune responses. Symptoms of adrenal insufficiency include fatigue, low blood pressure, hypoglycemia, muscle weakness, and reduced resistance to infections. In addition, it has psychological effects such as depression, anxiety, and reduced resilience to stress. The cycle of chronic stress and cortisol deficiency weakens the ability to cope with external stress. 

Two causes of adrenal insufficiency, also known as hypocortisolism, are modeled in this paper. One is primary adrenal insufficiency (adrenal disorder), and the other is secondary adrenal insufficiency, that is, hypothalamic-pituitary (HP) disorder. Understanding the effects of stress-related disorders can lead to early detection and targeted therapies. Therefore, it is crucial to address the effects of AI to reduce its physical, psychological, and social impacts by advancing our understanding of stress-related health disorders.

The HPA axis shown in Fig.~\ref{fig:HPA-axis} illustrates the physiological response to stress. When one is exposed to stress, the hypothalamus releases corticotropin-releasing hormone (CRH), which in turn stimulates the pituitary gland to secrete adrenocorticotropic hormone (ACTH). ACTH then travels through the bloodstream to the adrenal glands, where it stimulates the release of cortisol. The HPA axis has negative feedback loops that inhibit elevated CRH and ACTH production.

We now present a brief literature review on the subject of modelling and analysis of the HPA axis and adrenal insufficiency.

\subsection{Literature Review}
\label{Sec:literture rev}

\begin{figure*}[t]
    \centering
    \includegraphics[width=\textwidth]{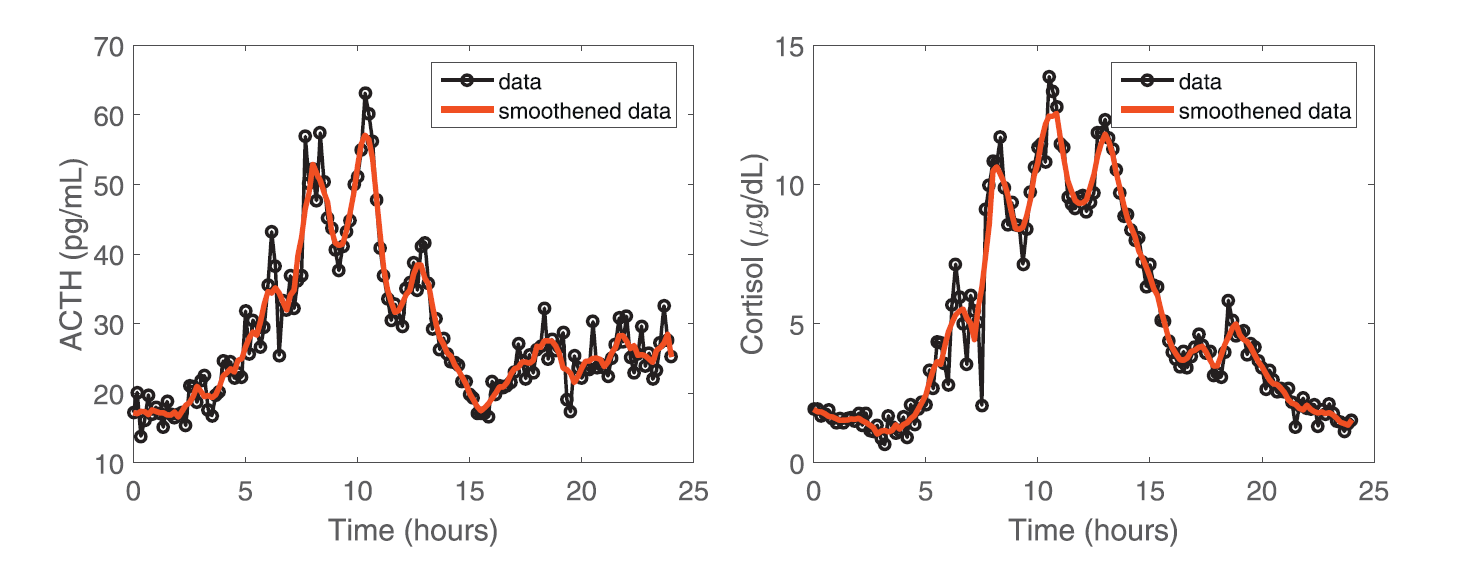} 
    \caption{The raw data of ACTH and Cortisol concentration over a day as reported by~\cite{BANGSGAARD}. (Applied for permission to reproduce the figure)}
    \label{fig:Rawdata}
\end{figure*}

In~\cite{CRH}, it was determined that the normal ranges for CRH in humans are 2--28 pg/mL. The normal range for ACTH is 5--50 pg/mL, and for cortisol is 3--25 µg/dL, as given by ~\cite{CortisolACTH2023}. Likewise, Fig.~\ref{fig:Rawdata} refers to the healthy ranges of ACTH and cortisol reported by~\cite{BANGSGAARD}. As noted by~\cite{SPEER2019100180}, the hormone ranges can vary depending on age, gender, stress level, history of trauma, weight, smoking status, alcohol or substance abuse, duration of sleep, menstrual cycle, etc. Elevated cortisol levels lead to hypercortisolism, while low levels result in adrenal insufficiency. 

Our paper focuses on adrenal insufficiency, which can be diagnosed by blood, saliva, or urine tests. Although there is no permanent cure for adrenal insufficiency, the condition is generally managed with appropriate medication and a healthy lifestyle. 

Our objective is to devise treatment strategies for patients suffering from adrenal insufficiency using a model-based approach. In what follows, we review existing HPA axis models, perform a comparative analysis, and point to the models that we will use in our study.

\cite{Vadim} developed a model of the HPA axis that contains two nonlinear factors which are the limit of gland size and the exclusion of non-physiological negative hormone concentrations. \cite{GUDMANDHOEYER201423} identified three potential biomarkers of depression through parameter analysis in their patient-specific modeling. \cite{Karin} explored the effects of long-term stress on the mass of corticotrophs and adrenal glands. Corticotrophs are present in the pituitary gland, which secretes ACTH, while the adrenal gland secretes cortisol. This leads to hormonal imbalances. \cite{Sriram} analysed normal as well as depressed and PTSD-affected individuals based on negative cortisol feedback. Complementing this approach,~\cite{Gupta} focused on physiological short-term and long-term stress responses on the HPA axis. \cite{Markovic} created a mathematical model to understand one's response to acute and chronic stress. They applied bifurcation analysis to predict the change in HPA axis, circadian, and ultradian rhythms under chronic stress.

Based on the concept of targeted therapy, a model predictive control approach was proposed in~\cite{BenZvi2009} to correct adrenal insufficiency using simulation-based strategies to optimize treatment protocols.~\cite{Ankush} presented explicit observer-based nonlinear predictive control to rectify HPA axis dysfunctions. They proposed a method to estimate hormone levels and stress based on measured ACTH levels. Further advancing the application of HPA axis models,~\cite{8419406} proposed a particle filter algorithm to enable real-time prediction and adjustment based on incoming data of nonlinearity, non-Gaussian behavior, stochastic parameters, and process noise.
\begin{figure}[t!]
\centerline{\includegraphics[width=1.25\textwidth]{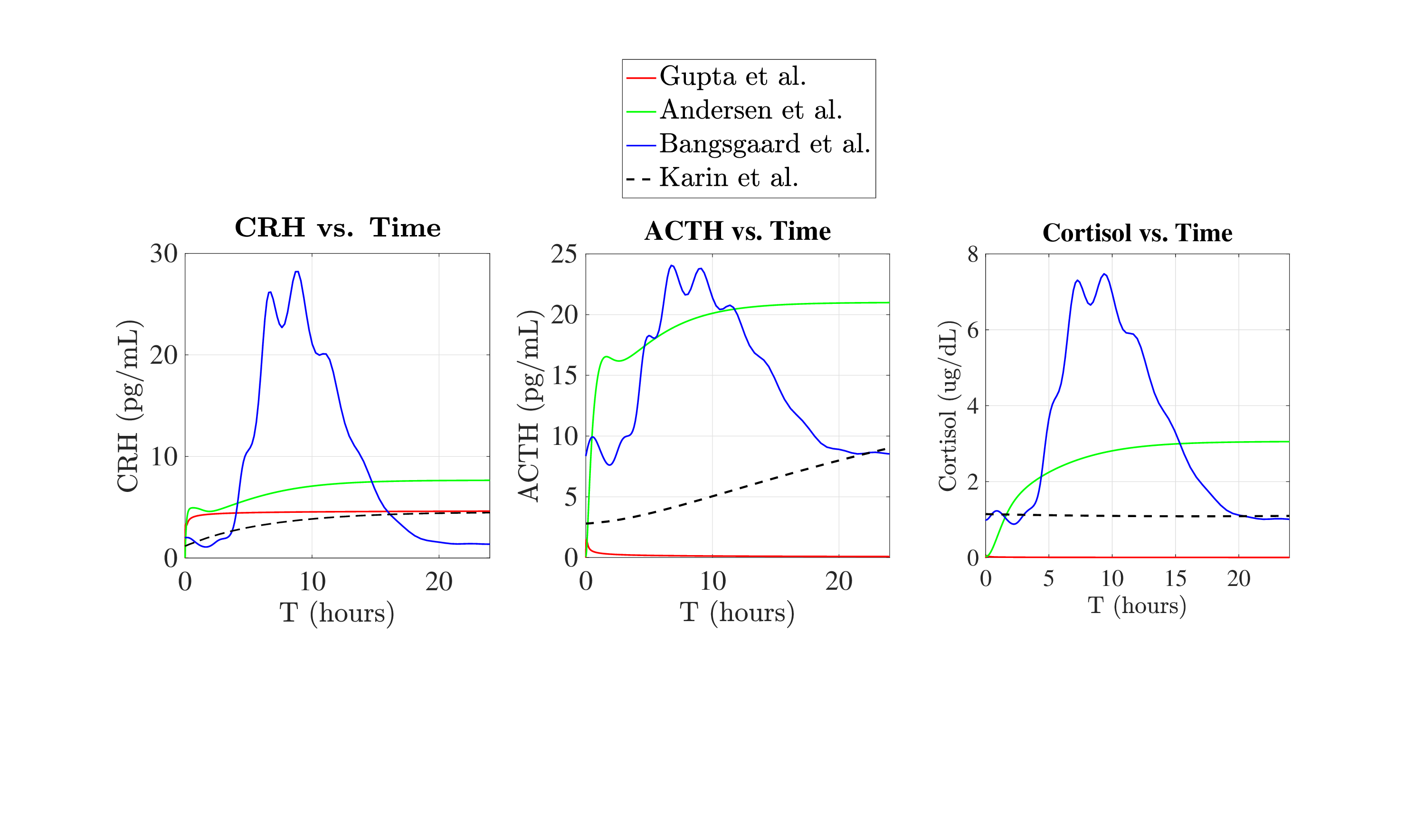}}
\caption{A comparative analysis of different HPA axis models demonstrated from literature. }
\label{fig:comparing three plots}
\end{figure}
~\cite{NIYOUSHA} reviewed five existing HPA axis models. They selected the model of ~\cite{ANDERSEN} and calibrated it based on empirical data from 17 healthy individuals and adjusted the model parameters, which reduced the mean absolute percent error by 71\% compared with available data. The model of ~\cite{ANDERSEN} was further explored by~\cite{BANGSGAARD}, incorporating patient-specific parameter estimation and accurately modeling circadian and ultradian rhythms. 

In our paper, we modify the model presented by~\cite{Karin}, in which we also incorporate the concept of circadian rhythm from~\cite{BANGSGAARD}. The result of combining the components of above models is a novel model that we use to generate treatment strategies for two conditions of adrenal insufficiency.

\subsection{Comparative Analysis}
In this section, we perform a comparative analysis of different HPA axis models from the literature. We used MATLAB Simulink to compare models proposed in~\cite{Gupta}, ~\cite{ANDERSEN},~\cite{NIYOUSHA}, ~\cite{BANGSGAARD}, and ~\cite{Karin}. We compared these models based on the following criteria: (i) the inclusion of circadian rhythm as an endogenous function of time, (ii) hormone concentration ranges secreted by the hypothalamus, pituitary, and adrenal glands, and (iii) the models' ability to simulate the hormone levels that correspond to two adrenal insufficiency conditions. A comparison of the plots from the models analyzed in this section has been carried out, which is demonstrated in Fig.~\ref{fig:comparing three plots}.

\cite{Gupta} and~\cite{Karin} did not consider endogenous circadian rhythm in their models. Hence, their plots for corresponding HPA axis hormones do not change throughout the day. \cite{Gupta} modeled the stress response of the HPA axis, whereas our objective was to construct a model that can simulate two types of AI. Therefore, while experimenting with their model, we set the stress input to zero. Although the model of~\cite{Gupta} was developed for a 10-hour window, we have extended the simulation to 24 hours to be consistent with the other models.

In contrast,~\cite{ANDERSEN} considered circadian rhythm as external input in the model but ultradian oscillations as an internal factor influencing the HPA axis. However, the hormone concentrations do not vary with time. \cite{NIYOUSHA} added a third-order function of time to the equation of cortisol of \cite{ANDERSEN} to simulate circadian oscillations as an exogenous function of time. We have not succeeded in obtaining physiologically plausible plots with~\cite{NIYOUSHA} model implemented in Simulink.  


\cite{BANGSGAARD} included the endogenous circadian rhythm in their model. They simulated healthy and hypercortisolemic individuals. We experimented with their model to produce results for two hypocortisolemic cases, fine-tuning some of the parameter values. Parameter values were chosen so that our model to fit patient data from literature. We observed that the CRH levels are too low and not in the expected range. The reason behind this is the second term in~\eqref{tag1} that approaches zero. As a result, the CRH levels do not vary as expected with the change in cortisol level. However, cortisol and ACTH levels are within the expected ranges. 

In our proposed model, we used the model of~\cite{Karin} as a baseline. We have modified this baseline model and also incorporated the endogenous circadian rhythm from~\cite{BANGSGAARD}. This modified model produces physiologically plausible results for modeling two hypocortisolemic conditions that are discussed in detail in this paper. Furthermore, our proposed model captures the known physiology of the HPA axis.


\subsection{Paper's Contribution}
Our paper's contributions include:
\begin{enumerate}
    \item Incorporation of circadian and ultradian oscillations in the proposed new model of the HPA axis.  \item Simulation of two distinct hypocortisolemic conditions: (a) primary adrenal insufficiency and (b) secondary adrenal insufficiency. 
     \item Introduction of an input variable $u$ to model cortisol replacement treatment strategies for hypocortisolemic patients. 
      \item Validation using published patient data with primary adrenal insufficiency.
      \item Simulation of theoretical CRH concentration to obtain further insight into HPA axis functioning. 
       \item Comparative analysis of different HPA axis models from the literature against the proposed HPA axis model in this paper.
\end{enumerate}

\section{The HPA axis model of controlled group}
\label{sec:Discussion on existing HPA axis models}

In this section, we discuss an HPA axis model proposed by~\cite{BANGSGAARD}. We use it to model healthy individuals in our paper.
The normal range of hormones for healthy individuals is taken from~\cite{CortisolACTH2023}
 and ~\cite{CRH}. We note that the model of~\cite{BANGSGAARD} is an extension of the model of~\cite{ANDERSEN}, where circadian and ultradian rhythm were added as endogenous functions. \cite{BANGSGAARD} have used their model to generate data for the controlled group. This model also replicates the empirical data of normal individuals presented in~\cite{Faghih2015}. The model of~\cite{BANGSGAARD} has the form:

\begin{align}
\frac{dx_1}{dt} &= a_0 + C(t) \frac{a_1}{1 + a_2 x_3^2} \frac{x_1}{\mu + x_1} - \omega_1 x_1, \label{tag1} \\
\frac{dx_2}{dt} &= \frac{a_3 x_1}{1 + a_4 x_3} - \omega_2 x_2, \label{tag2} \\
\frac{dx_3}{dt} &= a_5 x_2^2 - \omega_3 x_3, \label{tag3} \\
C(t) &= N_c \left( \frac{t_m^k}{t_m^k + \alpha^k} \times\frac{(T - t_m)^l}{(T - t_m)^l + \beta^l} + \epsilon \right), \label{tag4}
\end{align}
 where $x_1$, $x_2$, and $x_3$ are concentrations of CRH, ACTH, and cortisol, respectively. The function $C(t)$ given by~\eqref{tag4} models the circadian rhythm, where $t \in [0, T]$, with $T = 1440$ minutes $= 1$ day. Note that $C(t)$ is bounded within the range $[\epsilon, 1]$, where $\epsilon$ is a small positive number which is a design parameter. The value of $t_m$ is calculated as $t_m$ = $(t -\delta) \mod T$, where $\delta$ is the time shift characterizing the circadian rhythm, and $N_c$ is the normalization constant, which is the model's design parameter.  
 
 \begin{table}[h!]
\centering
\caption{Model parameters of controlled subjects from~\cite{BANGSGAARD}.}
\begin{tabular}{@{}clll@{}}
\hline
\textbf{No.} & \textbf{Parameter} & \textbf{Value} & \textbf{Unit} \\ 
\hline
1  & $a_0$    & $3.9031 \times 10^{-4}$ & pg/(mL$\times$min) \\ 
2  & $a_1$    & $0.75 \times 10^{13}$ & pg/(mL$\times$min) \\ 
3  & $a_2$    & $1.7809 \times 10^9$    & (dL/$\mu$g)$^2$ \\ 
4  & $\mu$    & $583$    & pg/mL \\ 
5  & $a_3$    & $2.2803 \times 10^4$    & min$^{-1}$ \\ 
6  & $a_4$    & $1.7745 \times 10^5$    & dL/$\mu$g \\ 
7  & $a_5$    & $4.6170 \times 10^{-4}$ & $\mu$g/dL \\ 
8  & $\omega_1$ & $0.0337$              & min$^{-1}$ \\ 
9  & $\omega_2$ & $0.0205$              & min$^{-1}$ \\ 
10 & $\omega_3$ & $0.02$              & min$^{-1}$ \\ 
11 & $\delta$  & $83.8$                 & min \\ 
12 & $\alpha$  & $300$                 & min \\ 
13 & $k$       & $5$                   & -- \\ 
14 & $\beta$   & $950$                 & min \\ 
15 & $l$       & $6$                   & -- \\ 
16 & $\epsilon$ & $0.01$               & -- \\ 
17 & $N_c$     & $0.5217$              & -- \\ 
\hline
\end{tabular}
\label{tab:parameters1}
\end{table}

The parameter $k$ is the steepness of the increasing Hill function at $t_m = \alpha$, while $l$ is the steepness of the decreasing Hill function at $t_m = \beta$, where $\alpha$ and $\beta$ are two design constants. The parameter $a_0$ corresponds to a low level of CRH secretion, while $a_1$, $a_3$, and $a_5$ control the rate of change in the production of CRH, ACTH, and cortisol, respectively. Additionally, $a_2$ inhibits the production of CRH, whereas $a_4$ inhibits the production of ACTH. The parameters $\omega_1$, $\omega_2$, and $\omega_3$ are the degradation constants of the corresponding hormones, and $\mu$ represents the Michaelis-Menten constant. 

The cortisol levels in the morning around 8:00 am should be normally in the range of 8--10 $\mu$g/dL. While reproducing the plots by~\cite{BANGSGAARD}, we observed that the cortisol barely reached the optimum level. Thus, we have fine-tuned the model parameter values of~\cite{BANGSGAARD} for the model to attain the desired levels. The modified values are listed in Table~\ref{tab:parameters1}. 

\section{Modeling Adrenal Insufficiency (AI)}
In this section, we propose a new HPA axis model for simulating two abnormal conditions: hypothalamus-pituitary disease and primary adrenal disease. We used the model of~\cite{Karin} as a starting point for our proposed model. A key advantage of our model is the inclusion of the circadian rhythm as an endogenous function of time. We simulated the hypocortisolemic cases using the equations given by ~\cite{Karin}, but the results were not satisfactory. The modified equations are discussed later in this section.
 The circadian rhythm, along with feedback loops, plays an essential role in the functioning of the HPA axis as noted in~\cite{Circadian}. Both synchronize the release of cortisol with diurnal variations. The oscillatory behavior of the circadian function is due to negative feedback loops from cortisol to CRH and ACTH. The oscillatory behavior can be disturbed by an external noise. The interaction between the circadian rhythm and the feedback loop ensures stable functioning of the HPA axis when our body is exposed to external stress or internal disturbances. The body's circadian and ultradian pulses manage the resource allocation, followed by the utilization of energy to help smooth the functioning of all the organs, thereby maintaining homeostasis. 

\begin{figure}[!h]
\centerline{\includegraphics[width=0.5\textwidth]{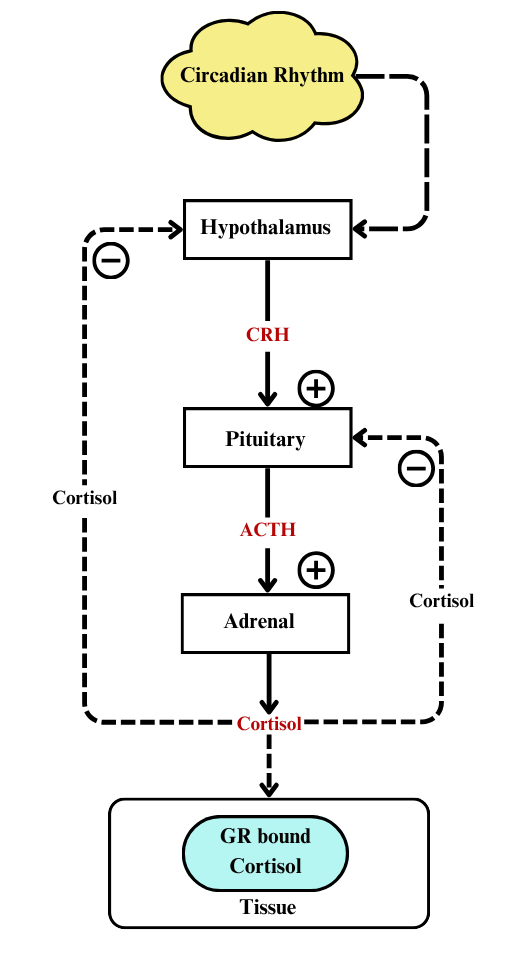}}
\caption{Block diagram of the proposed HPA axis model.}
\label{fig:HPA-axis-block diagram}
\end{figure}

The HPA axis model proposed by~\cite{Karin} consists of a feedback mechanism that takes into account glucocorticoid receptors (GR) and mineralocorticoid receptors (MR). Cortisol binds to GR and provides negative feedback to the hypothalamus and pituitary gland, as shown in Fig.~\ref{fig:HPA-axis-block diagram}. In our proposed model, we incorporate the circadian rhythm function, $C(t)$ given by~\eqref{tag4}. The resulting model has the form:
 
\begin{align}
\frac{dx_1}{dt} &= K_0 + g_1(x_3) C(t) K_1 x_1 - \omega_1x_1 , \label{tag5} \\
\frac{dx_2}{dt} &= K_2 g_1(x_3)x_1 - \omega_2 x_2,\label{tag6} \\
\frac{dx_3}{dt} &= K_3 x_2 - \omega_3x_3 + {u},\label{tag7} \\
g_1(x_3) &= M(x_3)G(x_3),\label{tag8} \\
M(x_3) &= \frac{1}{x_3},\label{tag9} \\
G(x_3) &= \frac{1}{1 + \left( \frac{x_3}{K_{GR}} \right)^n},\label{tag10}
\end{align}
where $x_1$, $x_2$, and $x_3$ are concentrations of CRH, ACTH, and cortisol, respectively. We introduced an input variable $u$ to model the treatment in~\eqref{tag7}. In~\eqref{tag5}, $K_0$ is a small positive constant to ensure the non-negative level of CRH. The Hill function $g_1(x_3)$ models the cortisol feedback to both CRH and ACTH, $M(x_3)$ and $G(x_3)$ are also Hill functions that model how MR and GR bind with cortisol. The GR and MR receptors become active as cortisol increases. The parameter $K_{GR}$ denotes the amount of cortisol required to decrease the activity of GR receptors to half. Furthermore, $K_1$, $K_2$, and $K_3$ are production constants, and $\omega_1$, $\omega_2$, and $\omega_3$ are hormone degradation constants of corresponding HPA axis hormones. In~\eqref{tag10}, $n$ is a tuning parameter. The parameter values of the proposed model are given in Table~\ref{tab:parameters2}. 

\begin{table}[h!]
\centering
\caption{Model parameters for an adrenal insufficient patient.}
\label{tab:parameters2}
\begin{tabular}{@{}cllll@{}} 
\hline
\textbf{No.} & \textbf{Parameter} & \textbf{Primary AI} & \textbf{Secondary AI} & \textbf{Unit} \\ 
\hline
1  & $K_0$    & 0.008   & 0.02 & pg/(mL$\times$min) \\ 
2  & $K_1$    & 0.0085  & 0.02 & pg/(mL$\times$min) \\ 
3  & $K_2$    & 0.005 & 0.003 & min$^{-1}$ \\ 
4  & $K_3$    & 0.000009 & 0.0009 & min$^{-1}$ \\ 
5  & $\omega_1$    & 0.0009  & 0.009 & min$^{-1}$ \\ 
6  & $\omega_2$    & 0.0019  & 0.001 & min$^{-1}$ \\ 
7  & $\omega_3$    & 0.003 & 0.002 & min$^{-1}$ \\  
8  & $K_{GR}$    & 4 & 4 & -- \\ 
9  & $n$    & 3 & 3 & -- \\ 
\hline
\end{tabular}
\label{tab:parameters2}
\end{table}

 We use our model to simulate two different adrenal insufficiency conditions with different parameter values for each case. The two types of adrenal insufficiency are primary and secondary AI. Primary AI is characterized by elevated levels of CRH and ACTH due to the failure of the adrenal glands to produce a sufficient amount of cortisol. Whereas in secondary AI, the hypothalamus and pituitary glands are unable to produce adequate amounts of CRH and ACTH hormones, respectively. As a result, the adrenal glands cannot generate the necessary cortisol levels. This disrupts the negative feedback mechanism to the hypothalamus and pituitary. To regulate these HPA axis disorders, we used an open-loop cortisol replacement medication strategy. 

\begin{figure}[t!]
\noindent\makebox[\textwidth][c]{
\includegraphics[width=\paperwidth,height=\paperheight,keepaspectratio]{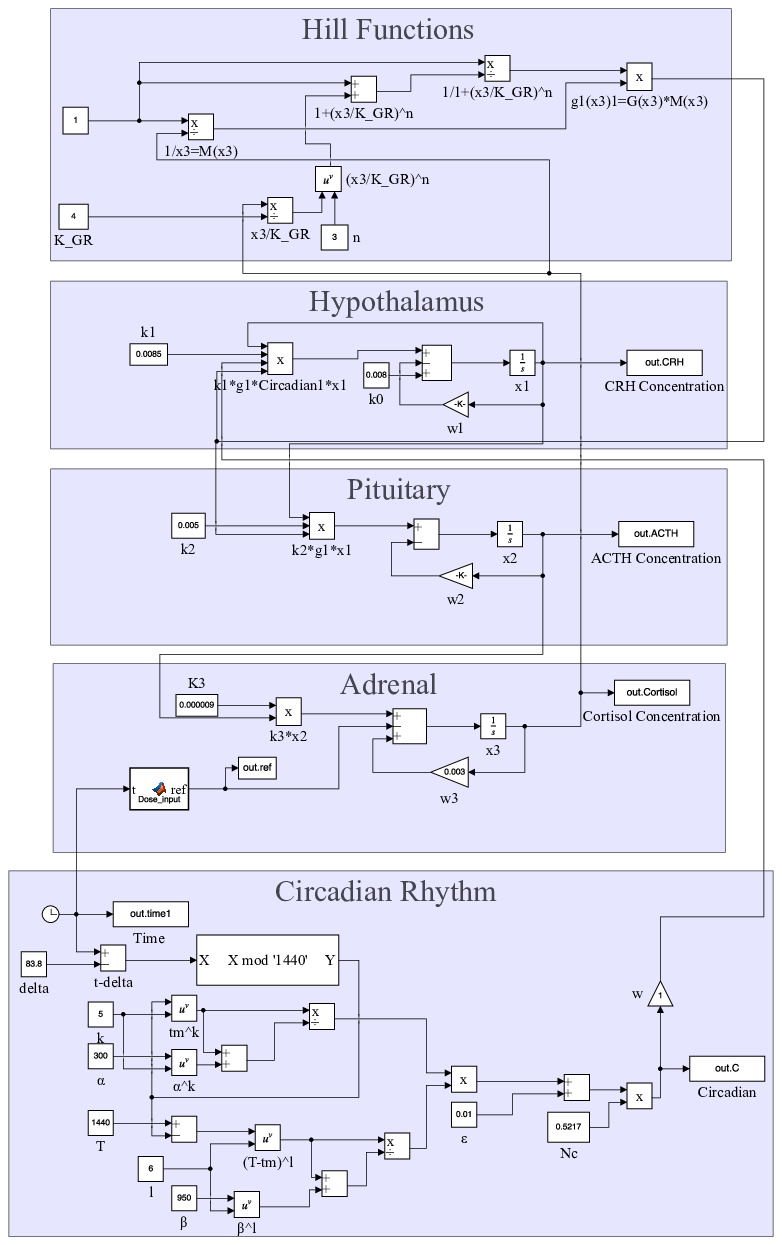}
}
\caption{Simulink diagram of the proposed HPA axis model.}
\label{fig:For_Simulation_new}
\end{figure}

A simplified block diagram of the HPA axis is illustrated in~Fig.~\ref{fig:HPA-axis-block diagram}. This figure shows sequential hormonal signaling from the hypothalamus to the adrenal glands, with circadian rhythm and negative feedback loops. The block diagram in~Fig~\ref{fig:HPA-axis-block diagram} complements the Simulink model shown in Fig.~\ref{fig:For_Simulation_new} by providing a conceptual understanding of the dynamics of the HPA axis. The detailed Simulink model of the HPA axis is divided into five sections: hill functions, circadian rhythm, hypothalamus, pituitary, and adrenal glands. Each section consists of interconnected mathematical blocks that represent the differential equations~\eqref{tag4},~\eqref{tag5},~\eqref{tag6},~\eqref{tag7},~\eqref{tag8},~\eqref{tag9},~\eqref{tag10} that control hormonal interactions.

\section{Treatment strategy}
In this section, we discuss about cortisol replacement strategy for individuals suffering from hypocortisolemic disorders. Cortisol concentration varies throughout the day according to the circadian rhythm. The term cortisol-awakening response (CAR) is characterized by elevated cortisol levels in the morning, as noted by \cite{K}. ~\cite{NIYOUSHA} and~\cite{Carroll} suggested that the cortisol levels are typically low from 8:00pm to 4:00am, peak between 4:00am and 12:00pm, and then decline back to minimal levels from 12:00pm until 8:00pm. As recommended by the Endocrine Society, noted by~\cite{Bornstein}, cortisol replacement therapy typically involves taking 15--25 mg of hydrocortisone in two or three divided doses throughout the day. We use MATLAB Simulink software in our modeling. In our simulation, we model cortisol replacement therapy administered three times a day for primary and secondary AI, respectively. 

\subsection{Primary Adrenal Insufficiency}
\begin{figure}[h]
\centering
\includegraphics[width=\textwidth]{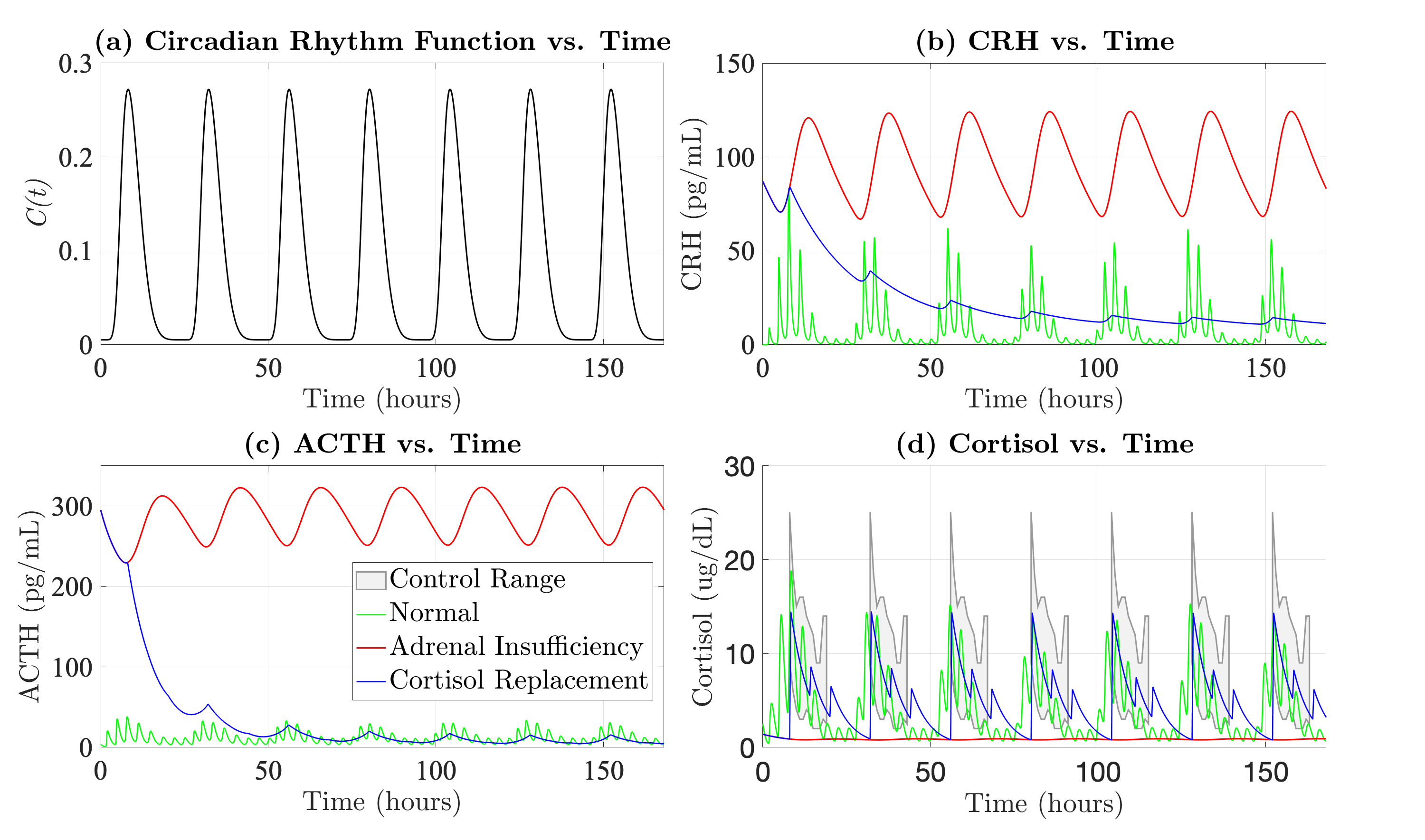}
\caption{Comparison of hormone levels of the HPA axis for normal, primary adrenal insufficiency, and after cortisol replacement therapy for a week. The labels are the same as in sub-figure (c) of Fig.~\ref{fig:Graph-adrenal-case}.}
\label{fig:Graph-adrenal-case}
\end{figure}

\begin{figure}[H]
\centering
\includegraphics[width=\textwidth]{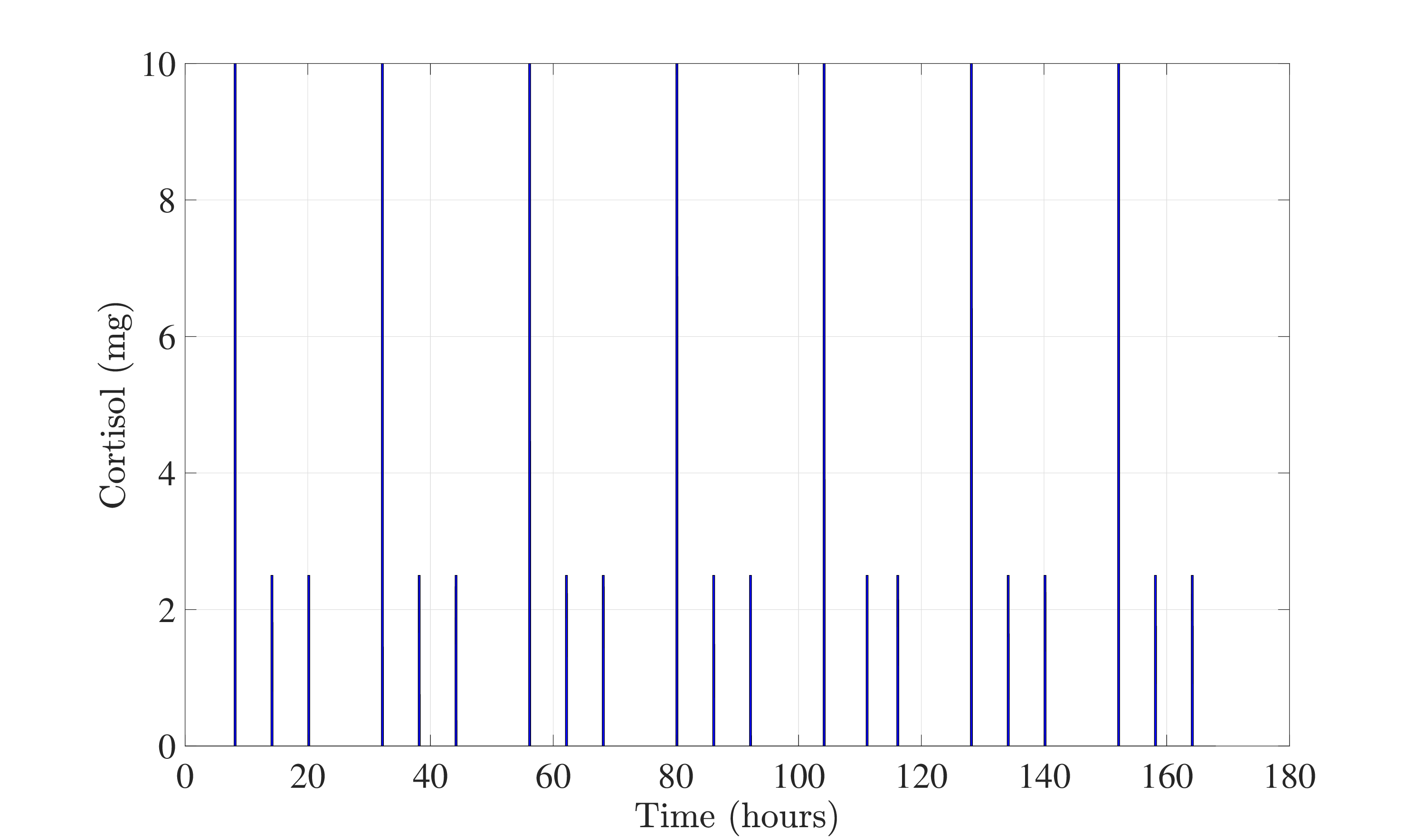}
\caption{Cortisol replacement therapy for primary adrenal insufficiency. A total dosage of 21~mg in three divided doses of hydrocortisone was administered as follows: 14~mg, 3.5~mg, and 3.5~mg at 8:00~am, 2:00~pm, and 8:00~pm, respectively, for one week.}
\label{fig:Medication_Adrenal}
\end{figure}

Primary adrenal insufficiency is also known as Addison's disease. We have compared the simulated hormone levels of a normal individual with the patient suffering from adrenal insufficiency, with and without medication, which is illustrated in~Fig.~\ref{fig:Graph-adrenal-case}. In this condition, the adrenal gland does not function or is damaged. Therefore, the glands cannot produce the required amount of cortisol. As a result, ACTH and CRH increase and the levels remain high, seeking more cortisol. In this case, we apply 21 mg cortisol externally in three doses 14~mg, 3.5~mg, and again 3.5~mg each day for a week, as shown in Fig.~\ref{fig:Medication_Adrenal}. After medication, the cortisol level increases and reaches the steady state as shown in Fig.~\ref{fig:Graph-adrenal-case}{(d)}. Initially, the ACTH concentration drops drastically to the normal range within 24 hours. Thereafter, the ACTH levels gradually decrease over the period as depicted in~Fig~\ref{fig:Graph-adrenal-case}{(c)}. CRH concentration also decreases to a certain value, a little higher than the healthy range, which stabilizes to a normal level as shown in~Fig~\ref{fig:Graph-adrenal-case}{(b)}.
The levels of CRH and ACTH reduce after the medication due to the negative feedback from cortisol. 

\begin{figure}[H]
\centering
\includegraphics[width=\textwidth]{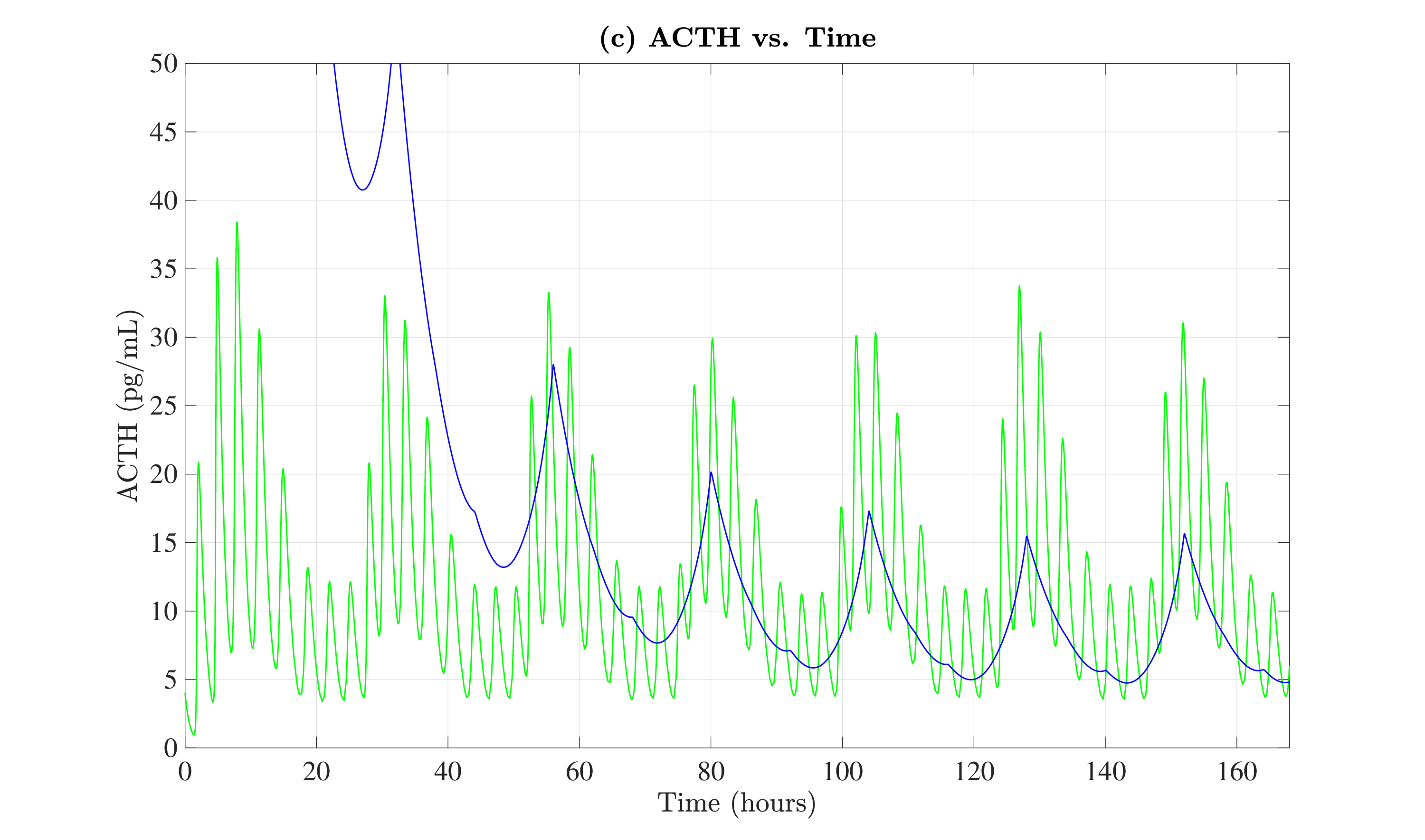}
\caption{Detailed view of the levels of ACTH for primary adrenal insufficiency from sub-figure (c) of Fig.~\ref{fig:Graph-adrenal-case}.}
\label{fig:Detailed view}
\end{figure}

\subsection{Secondary Adrenal Insufficiency}
\begin{figure*}[t] 
\centering
\includegraphics[width=\textwidth]{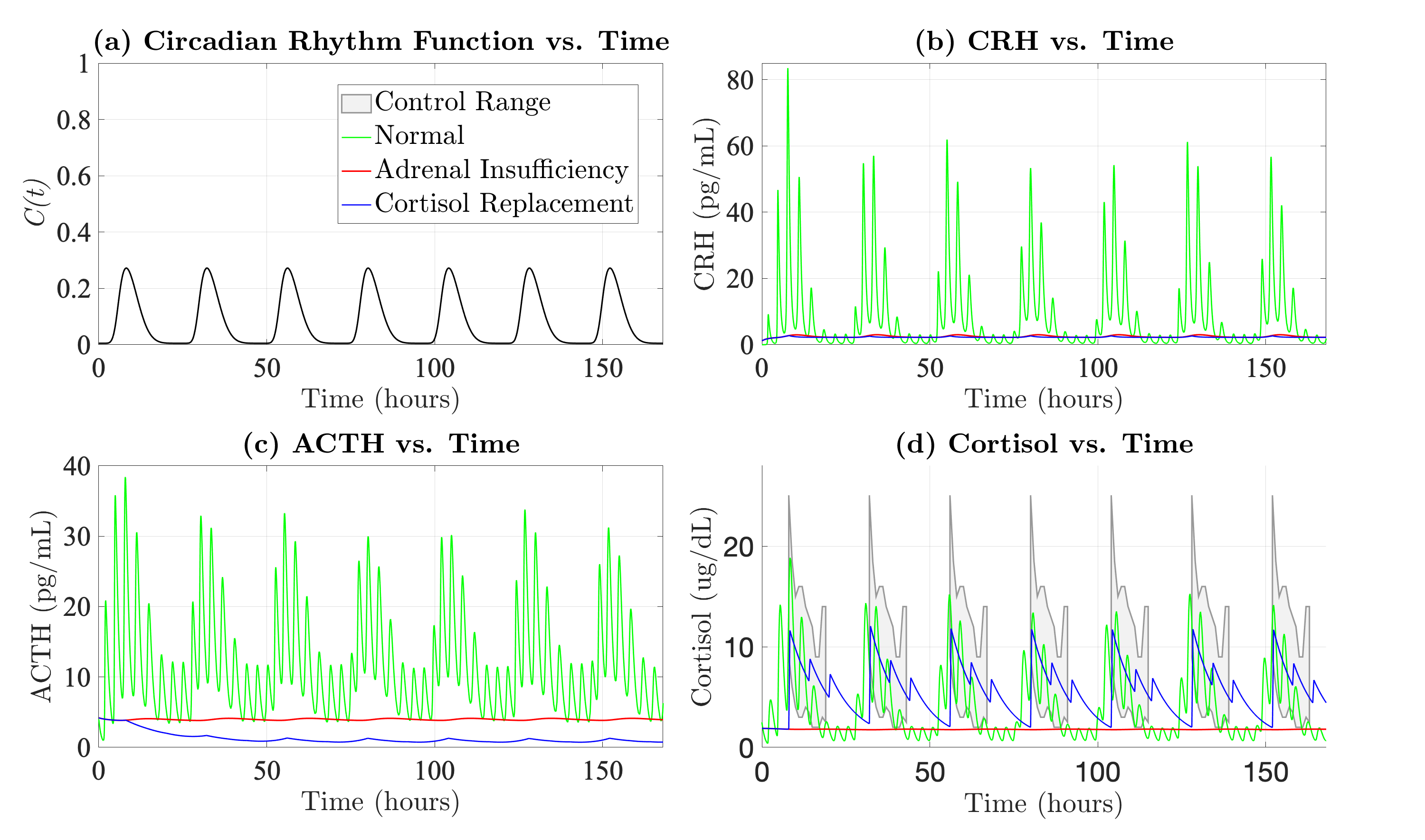} %
\caption{Comparison of hormone levels of the HPA axis for normal, secondary adrenal insufficiency, and after cortisol replacement therapy over a week. The legends are the same as in sub-figure (a). }
\label{fig:Graph-HP-case}
\end{figure*}

\begin{figure}[H]
\centering
\includegraphics[width=\textwidth]{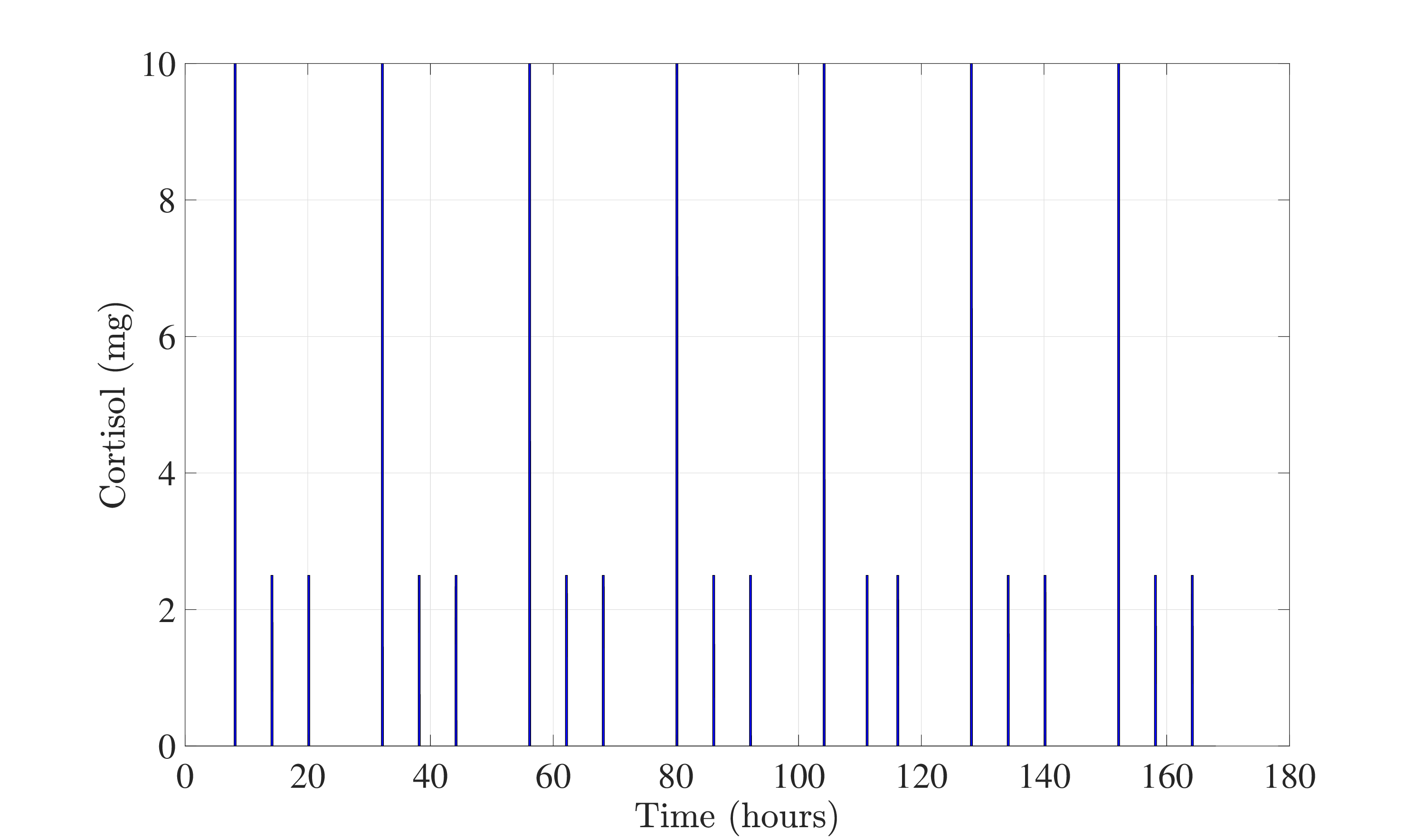}
\caption{Cortisol replacement therapy for secondary adrenal insufficiency. Total dosage of 15~mg in three divided doses of 10~mg, 2.5~mg, and 2.5~mg at 8:00~am, 2:00~pm, and 8:00~pm, respectively, for one week.}
\label{fig:Medication_HP}
\end{figure}

In hypothalamus-pituitary disorder, also known as secondary adrenal insufficiency, the hypothalamus and pituitary gland do not synthesize enough CRH and ACTH, respectively. As a result, the adrenal gland does not get the stimulus to secrete the necessary amount of cortisol. We have simulated this scenario, and the results are presented in Fig.~\ref{fig:Graph-HP-case}. We have compared the hormone levels of a healthy individual, one with a HP disorder, and one with HP disorder given cortisol replacement therapy for a week. From the graph, we can infer that the concentration of ACTH and CRH remains low, and therefore, the cortisol is also low. In our study, we have administered 15 mg of cortisol externally in the form of hydrocortisone for secondary adrenal insufficiency. The total dosage is divided into three doses: 10~mg, 2.5~mg, and again 2.5~mg per day, as shown in Fig.~\ref{fig:Medication_HP}. After administering hydrocortisone at 8:00 am, 2:00 pm, and 8:00 pm on a daily basis, the cortisol level increases to a healthy range as shown in Fig~\ref{fig:Graph-HP-case} {(d)}. The cortisol level saturates over a period. Whereas, ACTH and CRH remain low as shown in~Fig~\ref{fig:Graph-HP-case}{(c)} and~Fig~\ref{fig:Graph-HP-case}{(b)}, respectively. The amount of CRH and ACTH reduces because of the negative feedback from cortisol. So, as the cortisol level elevates after the medication, it signals the corresponding tissues to cease the secretion of CRH and ACTH. 

Individuals suffering from adrenal insufficiency (AI) require life long treatment with glucocorticoids. We simulated cases where various amounts of doses of hydrocortisone were administered. We arrived at a specific input doses for both primary and secondary AI through simulations. The parameters were refined through successive simulations based on sensitivity of the parameters. This process was repeated until we have generated curves that are consistent with known physiology. 
In primary AI the adrenal glands are not functioning and hence no cortisol is produced. This requires a higher cortisol replacement dose compared with the secondary AI where some cortisol is still being produced. While choosing the best input dosage, we factored in commonly available doses. The estimated dosage can be obtained by cutting the pills to make the dose smaller or larger as needed. However, this may result in either overdosing or underdosing. This underlines the necessity for research to develop a broader range of pills to ensure optimal medication doses.

\begin{figure}[H]
\centering
\includegraphics[width=\textwidth]{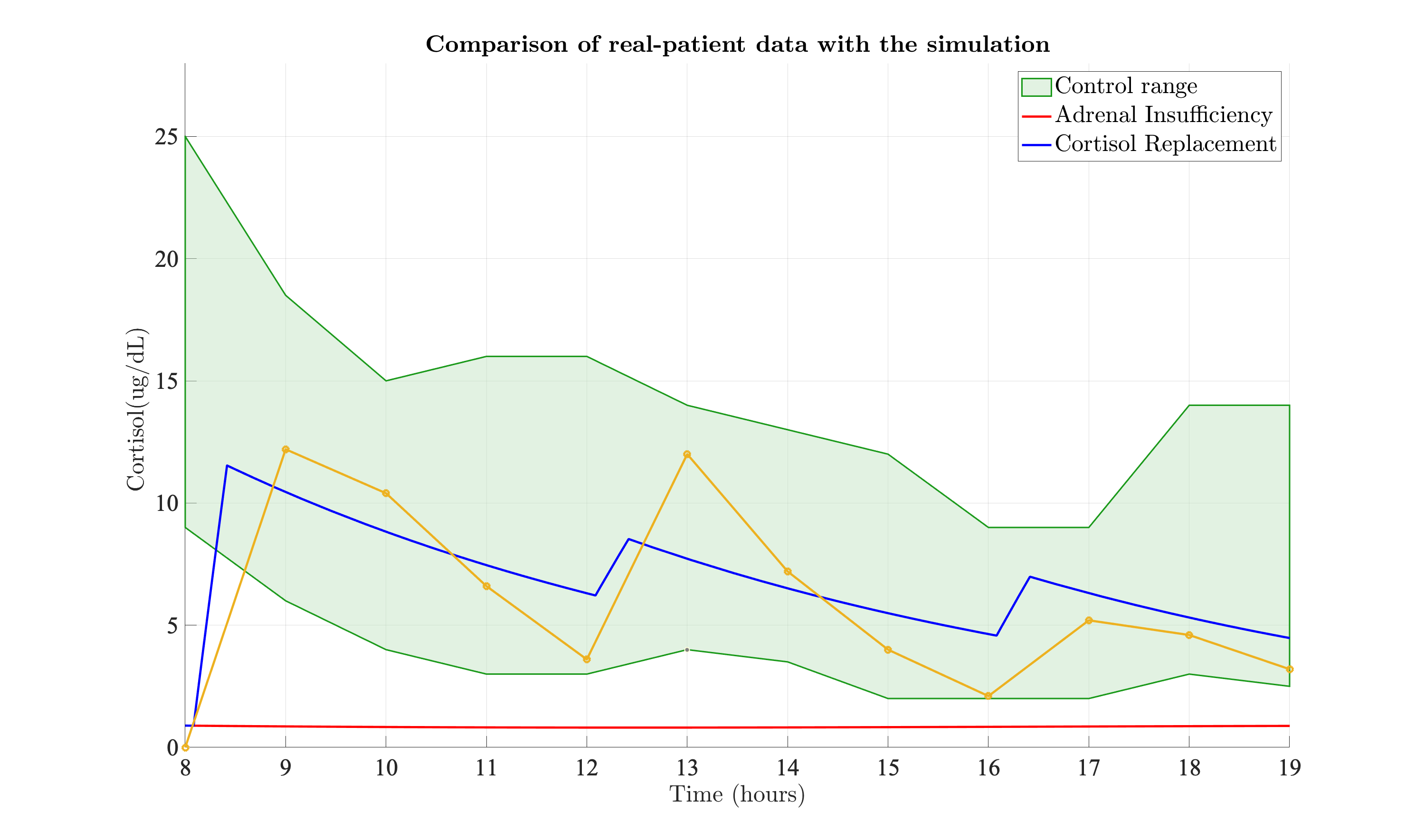}
\caption{Comparison of published patient data with AI being treated with cortisol replacement therapy (yellow line) from Rousseau et al. (2015), against the simulated patient with AI, using our model, being treated with our treatment strategy (blue line): same dosage of 17.4 mg ($9.4~\mathrm{mg/m^2/day}$) of cortisol in three divided doses, 11.6 mg, 2.9 mg and 2.9 mg at 8:05 am, 12:05 pm and 4:05 pm, respectively.}
\label{fig:Data validation}
\end{figure}

We tested our model against published patient data of primary AI being treated with cortisol replacement therapy. The data come from~\cite{Rousseau2015}.  We used our model to simulate the patient with primary AI (red line) shown in Fig.~\ref{fig:Data validation}. We then administered a medication strategy to this patient. We administered 17.4~mg ($9.4~\mathrm{mg/m^2/day}$), divided into three doses. We gave 11.6~mg at 8:05~am, 2.9~mg at 12:05~pm, and again 2.9~mg at 4:05~pm. This translates approximately to 66.67\%, 16.67\%, and 16.67\% of the entire dosage throughout the day. The patient undergoing this strategy is shown in Fig.~\ref{fig:Data validation} in blue. For comparison, we have taken in account Body Surface Area (BSA). Typically from literature, 

As can be seen from the figure, the medication strategy applied to the patient results in cortisol plasma in an appropriate region, shaded in green. We conclude that the above comparison validates our proposed model of the HPA axis for primary AI. However, the simulated CRH curves for both primary and secondary AI could not be validated because no clinical data is currently available. These simulations should provide insight on how CRH should respond for validation in future studies.

\section{Summary and future work}
\subsection{Summary}

In this paper, we reviewed the HPA axis models and conducted a comparative analysis of them. Comparison metrics include recognition of circadian rhythm as an endogenous function, HPA axis hormone ranges, and the functionality of the model to produce a hypocortisolemic condition. A novel mathematical model of the HPA axis is proposed to model two conditions of hypocortisolemic disorders. Adrenal insufficiency is a condition in which the adrenal glands are unable to produce a sufficient amount of cortisol. Two simulated hypocortisolemic cases are hypothalamic-pituitary disease and primary adrenal disease. Cortisol replacement treatment strategies are proposed and simlulated. After administering cortisol replacement therapy, cortisol levels return to the healthy range. Mathematical models of HPA axis disorders followed by treatment strategies are consistent with patient data found in~\cite{Rousseau2015}.

\subsection{Future work}

The next stage of this research is to investigate the closed-loop control strategies. They include model-predictive control, fuzzy control, model reference adaptive control, or machine learning approaches, which can act as virtual assistants for physicians. The obtained results can then be validated in a clinical setting to advance patient-specific strategies in neuro-endocrinology. Further studies could be directed towards mathematical modeling of the interconnection of HPA and the hypothalamus-pituitary thyroid (HPT) axis. Both axes interact and have some influence on each other as stated in~\cite{HPA-HPT, HPA-HPT1, HPA-HPT2}. This can lead to a more accurate modeling of stress-related disorders. Additionally, modeling of the HPA axis can be a foundation for the education and training of students, researchers, and medical professionals in endocrinology.

\bibliography{sn-bibliography}


\begin{thebibliography}{25}
\ifx \bisbn   \undefined \def \bisbn  #1{ISBN #1}\fi
\ifx \binits  \undefined \def \binits#1{#1}\fi
\ifx \bauthor  \undefined \def \bauthor#1{#1}\fi
\ifx \batitle  \undefined \def \batitle#1{#1}\fi
\ifx \bjtitle  \undefined \def \bjtitle#1{#1}\fi
\ifx \bvolume  \undefined \def \bvolume#1{\textbf{#1}}\fi
\ifx \byear  \undefined \def \byear#1{#1}\fi
\ifx \bissue  \undefined \def \bissue#1{#1}\fi
\ifx \bfpage  \undefined \def \bfpage#1{#1}\fi
\ifx \blpage  \undefined \def \blpage #1{#1}\fi
\ifx \burl  \undefined \def \burl#1{\textsf{#1}}\fi
\ifx \doiurl  \undefined \def \doiurl#1{\url{https://doi.org/#1}}\fi
\ifx \betal  \undefined \def \betal{\textit{et al.}}\fi
\ifx \binstitute  \undefined \def \binstitute#1{#1}\fi
\ifx \binstitutionaled  \undefined \def \binstitutionaled#1{#1}\fi
\ifx \bctitle  \undefined \def \bctitle#1{#1}\fi
\ifx \beditor  \undefined \def \beditor#1{#1}\fi
\ifx \bpublisher  \undefined \def \bpublisher#1{#1}\fi
\ifx \bbtitle  \undefined \def \bbtitle#1{#1}\fi
\ifx \bedition  \undefined \def \bedition#1{#1}\fi
\ifx \bseriesno  \undefined \def \bseriesno#1{#1}\fi
\ifx \blocation  \undefined \def \blocation#1{#1}\fi
\ifx \bsertitle  \undefined \def \bsertitle#1{#1}\fi
\ifx \bsnm \undefined \def \bsnm#1{#1}\fi
\ifx \bsuffix \undefined \def \bsuffix#1{#1}\fi
\ifx \bparticle \undefined \def \bparticle#1{#1}\fi
\ifx \barticle \undefined \def \barticle#1{#1}\fi
\bibcommenthead
\ifx \bconfdate \undefined \def \bconfdate #1{#1}\fi
\ifx \botherref \undefined \def \botherref #1{#1}\fi
\ifx \url \undefined \def \url#1{\textsf{#1}}\fi
\ifx \bchapter \undefined \def \bchapter#1{#1}\fi
\ifx \bbook \undefined \def \bbook#1{#1}\fi
\ifx \bcomment \undefined \def \bcomment#1{#1}\fi
\ifx \oauthor \undefined \def \oauthor#1{#1}\fi
\ifx \citeauthoryear \undefined \def \citeauthoryear#1{#1}\fi
\ifx \endbibitem  \undefined \def \endbibitem {}\fi
\ifx \bconflocation  \undefined \def \bconflocation#1{#1}\fi
\ifx \arxivurl  \undefined \def \arxivurl#1{\textsf{#1}}\fi
\csname PreBibitemsHook\endcsname

\bibitem[\protect\citeauthoryear{Androulakis}{2021}]{Circadian}
\begin{barticle}
\bauthor{\bsnm{Androulakis}, \binits{I.P.}}:
\batitle{Circadian rhythms and the {HPA} axis: A systems view}.
\bjtitle{WIREs Mechanisms of Disease}
\bvolume{13}(\bissue{4}),
\bfpage{1518}
(\byear{2021})
\doiurl{10.1002/wsbm.1518} .
\bcomment{Epub 2021 Jan 12}
\end{barticle}
\endbibitem

\bibitem[\protect\citeauthoryear{Aron}{2023}]{CortisolACTH2023}
\begin{botherref}
\oauthor{\bsnm{Aron}, \binits{D.}}:
Normal Values of Cortisol and {ACTH}.
Cushing's Syndrome and Adrenal Disease Foundation.
\url{https://csrf.net/doctors-answers/diagnosis-qa/normal-values-of-cortisol-and-acth/}
Accessed 2023-01-26
\end{botherref}
\endbibitem

\bibitem[\protect\citeauthoryear{Andersen et~al.}{2013}]{ANDERSEN}
\begin{barticle}
\bauthor{\bsnm{Andersen}, \binits{M.}},
\bauthor{\bsnm{Vinther}, \binits{F.}},
\bauthor{\bsnm{Ottesen}, \binits{J.T.}}:
\batitle{Mathematical modeling of the hypothalamic–pituitary–adrenal gland {(HPA)} axis, including hippocampal mechanisms}.
\bjtitle{Mathematical Biosciences}
\bvolume{246}(\bissue{1}),
\bfpage{122}--\blpage{138}
(\byear{2013})
\doiurl{10.1016/j.mbs.2013.08.010}
\end{barticle}
\endbibitem

\bibitem[\protect\citeauthoryear{Bornstein et~al.}{2016}]{Bornstein}
\begin{barticle}
\bauthor{\bsnm{Bornstein}, \binits{S.R.}},
\bauthor{\bsnm{Allolio}, \binits{B.}},
\bauthor{\bsnm{Arlt}, \binits{W.}},
\bauthor{\bsnm{Barthel}, \binits{A.}},
\bauthor{\bsnm{Don-Wauchope}, \binits{A.}},
\bauthor{\bsnm{Hammer}, \binits{G.D.}},
\bauthor{\bsnm{Husebye}, \binits{E.S.}},
\bauthor{\bsnm{Merke}, \binits{D.P.}},
\bauthor{\bsnm{Murad}, \binits{M.H.}},
\bauthor{\bsnm{Stratakis}, \binits{C.A.}},
\bauthor{\bsnm{Torpy}, \binits{D.J.}}:
\batitle{Diagnosis and treatment of primary adrenal insufficiency: An endocrine society clinical practice guideline}.
\bjtitle{Journal of Clinical Endocrinology and Metabolism}
\bvolume{101}(\bissue{2}),
\bfpage{364}--\blpage{389}
(\byear{2016})
\doiurl{10.1210/jc.2015-1710} .
\bcomment{Epub 2016 Jan 13}
\end{barticle}
\endbibitem

\bibitem[\protect\citeauthoryear{Bangsgaard and Ottesen}{2017}]{BANGSGAARD}
\begin{barticle}
\bauthor{\bsnm{Bangsgaard}, \binits{E.O.}},
\bauthor{\bsnm{Ottesen}, \binits{J.T.}}:
\batitle{Patient specific modeling of the {HPA} axis related to clinical diagnosis of depression}.
\bjtitle{Mathematical Biosciences}
\bvolume{287},
\bfpage{24}--\blpage{35}
(\byear{2017})
\doiurl{10.1016/j.mbs.2016.10.007} .
\bcomment{50th Anniversary Issue}
\end{barticle}
\endbibitem

\bibitem[\protect\citeauthoryear{Ben-Zvi et~al.}{2009}]{BenZvi2009}
\begin{barticle}
\bauthor{\bsnm{Ben-Zvi}, \binits{A.}},
\bauthor{\bsnm{Vernon}, \binits{S.D.}},
\bauthor{\bsnm{Broderick}, \binits{G.}}:
\batitle{Model-based therapeutic correction of hypothalamic-pituitary-adrenal axis dysfunction}.
\bjtitle{PLoS Computational Biology}
\bvolume{5}(\bissue{1}),
\bfpage{1000273}
(\byear{2009})
\doiurl{10.1371/journal.pcbi.1000273}
\end{barticle}
\endbibitem

\bibitem[\protect\citeauthoryear{Chakrabarty et~al.}{2014}]{Ankush}
\begin{bchapter}
\bauthor{\bsnm{Chakrabarty}, \binits{A.}},
\bauthor{\bsnm{Buzzard}, \binits{G.T.}},
\bauthor{\bsnm{Corless}, \binits{M.J.}},
\bauthor{\bsnm{Zak}, \binits{S.H.}},
\bauthor{\bsnm{Rundell}, \binits{A.E.}}:
\bctitle{Correcting hypothalamic-pituitary-adrenal axis dysfunction using observer-based explicit nonlinear model predictive control}.
In: \bbtitle{2014 36th Annual International Conference of the IEEE Engineering in Medicine and Biology Society},
pp. \bfpage{3426}--\blpage{3429}
(\byear{2014}).
\doiurl{10.1109/EMBC.2014.6944359}
\end{bchapter}
\endbibitem

\bibitem[\protect\citeauthoryear{Carroll et~al.}{2007}]{Carroll}
\begin{barticle}
\bauthor{\bsnm{Carroll}, \binits{B.J.}},
\bauthor{\bsnm{Cassidy}, \binits{F.}},
\bauthor{\bsnm{Naftolowitz}, \binits{D.}},
\bauthor{\bsnm{Tatham}, \binits{N.E.}},
\bauthor{\bsnm{Wilson}, \binits{W.H.}},
\bauthor{\bsnm{Iranmanesh}, \binits{A.}},
\bauthor{\bsnm{Liu}, \binits{P.Y.}},
\bauthor{\bsnm{Veldhuis}, \binits{J.D.}}:
\batitle{Pathophysiology of hypercortisolism in depression}.
\bjtitle{Acta Psychiatrica Scandinavica}
\bvolume{115}(\bissue{s433}),
\bfpage{90}--\blpage{103}
(\byear{2007})
\doiurl{10.1111/j.1600-0447.2007.00967.x}
{\href{https://arxiv.org/abs/https://onlinelibrary.wiley.com/doi/pdf/10.1111/j.1600-0447.2007.00967.x}{{https://onlinelibrary.wiley.com/doi/pdf/10.1111/j.1600-0447.2007.00967.x}}}
\end{barticle}
\endbibitem

\bibitem[\protect\citeauthoryear{Faghih et~al.}{2015}]{Faghih2015}
\begin{barticle}
\bauthor{\bsnm{Faghih}, \binits{R.T.}},
\bauthor{\bsnm{Dahleh}, \binits{M.A.}},
\bauthor{\bsnm{Adler}, \binits{G.K.}},
\bauthor{\bsnm{Klerman}, \binits{E.B.}},
\bauthor{\bsnm{Brown}, \binits{E.N.}}:
\batitle{Quantifying pituitary-adrenal dynamics and deconvolution of concurrent cortisol and adrenocorticotropic hormone data by compressed sensing}.
\bjtitle{IEEE Transactions on Biomedical Engineering}
\bvolume{62}(\bissue{10}),
\bfpage{2379}--\blpage{2388}
(\byear{2015})
\doiurl{10.1109/TBME.2015.2427745} .
\bcomment{Epub 2015 Apr 29}
\end{barticle}
\endbibitem

\bibitem[\protect\citeauthoryear{Figure-HPA-Axis}{2023}]{ZellxHPA2023}
\begin{botherref}
\oauthor{\bsnm{Figure-HPA-Axis}}:
Zellx HPA Axis Testing.
Pharmaceutical Networking.
Accessed: 2023-01-26
(2023).
\url{https://www.pharmaceutical-networking.com/zellx-hpa-axis-testing/}
\end{botherref}
\endbibitem

\bibitem[\protect\citeauthoryear{Gupta et~al.}{2006}]{Gupta}
\begin{bchapter}
\bauthor{\bsnm{Gupta}, \binits{S.}},
\bauthor{\bsnm{Aslakson}, \binits{E.}},
\bauthor{\bsnm{Vernon}, \binits{S.D.}}:
\bctitle{Identifying a central nervous system perturbation that explains peripheral hypocortisolism by modeling the {HPA} axis}.
In: \bbtitle{19th IEEE Symposium on Computer-Based Medical Systems (CBMS'06)},
pp. \bfpage{550}--\blpage{554}
(\byear{2006}).
\doiurl{10.1109/CBMS.2006.96}
\end{bchapter}
\endbibitem

\bibitem[\protect\citeauthoryear{Gudmand-Hoeyer et~al.}{2014}]{GUDMANDHOEYER201423}
\begin{barticle}
\bauthor{\bsnm{Gudmand-Hoeyer}, \binits{J.}},
\bauthor{\bsnm{Timmermann}, \binits{S.}},
\bauthor{\bsnm{Ottesen}, \binits{J.T.}}:
\batitle{Patient-specific modeling of the neuroendocrine {HPA}-axis and its relation to depression: Ultradian and circadian oscillations}.
\bjtitle{Mathematical Biosciences}
\bvolume{257},
\bfpage{23}--\blpage{32}
(\byear{2014})
\doiurl{10.1016/j.mbs.2014.07.013} .
\bcomment{Multiscale models and methods in biomedicine}
\end{barticle}
\endbibitem

\bibitem[\protect\citeauthoryear{Helmreich et~al.}{2005}]{HPA-HPT1}
\begin{barticle}
\bauthor{\bsnm{Helmreich}, \binits{D.L.}},
\bauthor{\bsnm{Parfitt}, \binits{D.B.}},
\bauthor{\bsnm{Lu}, \binits{X.Y.}},
\bauthor{\bsnm{Akil}, \binits{H.}},
\bauthor{\bsnm{Watson}, \binits{S.J.}}:
\batitle{Relation between the hypothalamic-pituitary-thyroid {(HPT)} axis and the hypothalamic-pituitary-adrenal {(HPA)} axis during repeated stress}.
\bjtitle{Neuroendocrinology}
\bvolume{81}(\bissue{3}),
\bfpage{183}--\blpage{192}
(\byear{2005})
\doiurl{10.1159/000087001} .
\bcomment{Epub ahead of print}
\end{barticle}
\endbibitem

\bibitem[\protect\citeauthoryear{Hosseinichimeh et~al.}{2015}]{NIYOUSHA}
\begin{barticle}
\bauthor{\bsnm{Hosseinichimeh}, \binits{N.}},
\bauthor{\bsnm{Rahmandad}, \binits{H.}},
\bauthor{\bsnm{Wittenborn}, \binits{A.K.}}:
\batitle{Modeling the hypothalamus–pituitary–adrenal axis: A review and extension}.
\bjtitle{Mathematical Biosciences}
\bvolume{268},
\bfpage{52}--\blpage{65}
(\byear{2015})
\doiurl{10.1016/j.mbs.2015.08.004}
\end{barticle}
\endbibitem

\bibitem[\protect\citeauthoryear{Karin et~al.}{2020}]{Karin}
\begin{barticle}
\bauthor{\bsnm{Karin}, \binits{O.}},
\bauthor{\bsnm{Raz}, \binits{M.}},
\bauthor{\bsnm{Tendler}, \binits{A.}},
\bauthor{\bsnm{Bar}, \binits{A.}},
\bauthor{\bsnm{Korem~Kohanim}, \binits{Y.}},
\bauthor{\bsnm{Milo}, \binits{T.}},
\bauthor{\bsnm{Alon}, \binits{U.}}:
\batitle{A new model for the {HPA} axis explains dysregulation of stress hormones on the timescale of weeks}.
\bjtitle{Molecular Systems Biology}
\bvolume{16}(\bissue{7}),
\bfpage{9510}
(\byear{2020})
\doiurl{10.15252/msb.20209510}
\end{barticle}
\endbibitem

\bibitem[\protect\citeauthoryear{Kyrylov et~al.}{2005}]{Vadim}
\begin{barticle}
\bauthor{\bsnm{Kyrylov}, \binits{V.}},
\bauthor{\bsnm{Severyanova}, \binits{L.A.}},
\bauthor{\bsnm{Vieira}, \binits{A.}}:
\batitle{Modeling robust oscillatory behavior of the hypothalamic-pituitary-adrenal axis}.
\bjtitle{IEEE Transactions on Biomedical Engineering}
\bvolume{52}(\bissue{12}),
\bfpage{1977}--\blpage{1983}
(\byear{2005})
\doiurl{10.1109/TBME.2005.857671}
\end{barticle}
\endbibitem

\bibitem[\protect\citeauthoryear{Kudielka and Wüst}{2010}]{K}
\begin{barticle}
\bauthor{\bsnm{Kudielka}, \binits{B.M.}},
\bauthor{\bsnm{Wüst}, \binits{S.}}:
\batitle{Human models in acute and chronic stress: Assessing determinants of individual hypothalamus–pituitary–adrenal axis activity and reactivity}.
\bjtitle{Stress}
\bvolume{13}(\bissue{1}),
\bfpage{1}--\blpage{14}
(\byear{2010})
\doiurl{10.3109/10253890902874913}
{\href{https://arxiv.org/abs/https://doi.org/10.3109/10253890902874913}{{https://doi.org/10.3109/10253890902874913}}}.
\bcomment{PMID: 20105052}
\end{barticle}
\endbibitem

\bibitem[\protect\citeauthoryear{Linton et~al.}{1987}]{CRH}
\begin{barticle}
\bauthor{\bsnm{Linton}, \binits{E.A.}},
\bauthor{\bsnm{McLean}, \binits{C.}},
\bauthor{\bsnm{Kruseman}, \binits{A.C.N.}},
\bauthor{\bsnm{Tilders}, \binits{F.J.}},
\bauthor{\bsnm{Veen}, \binits{E.A.V.}},
\bauthor{\bsnm{Lowry}, \binits{P.J.}}:
\batitle{Direct measurement of human plasma corticotropin-releasing hormone by "two-site" immunoradiometric assay}.
\bjtitle{Journal of Clinical Endocrinology and Metabolism}
\bvolume{64}(\bissue{5}),
\bfpage{1047}--\blpage{1053}
(\byear{1987})
\doiurl{10.1210/jcem-64-5-1047}
\end{barticle}
\endbibitem

\bibitem[\protect\citeauthoryear{Markovi{\'c} et~al.}{2011}]{Markovic}
\begin{barticle}
\bauthor{\bsnm{Markovi{\'c}}, \binits{V.M.}},
\bauthor{\bsnm{{\v{C}}upi{\'c}}, \binits{{\v{Z}}.}},
\bauthor{\bsnm{Vukojevi{\'c}}, \binits{V.}},
\bauthor{\bsnm{Kolar-Ani{\'c}}, \binits{L.}}:
\batitle{Predictive modeling of the hypothalamic-pituitary-adrenal {(HPA)} axis response to acute and chronic stress}.
\bjtitle{Endocrine Journal}
\bvolume{58}(\bissue{10}),
\bfpage{889}--\blpage{904}
(\byear{2011})
\doiurl{10.1507/endocrj.ej11-0037} .
\bcomment{Epub 2011 Aug 18}
\end{barticle}
\endbibitem

\bibitem[\protect\citeauthoryear{Mohammadbagheri et~al.}{2018}]{8419406}
\begin{bchapter}
\bauthor{\bsnm{Mohammadbagheri}, \binits{A.}},
\bauthor{\bsnm{Lillas}, \binits{C.}},
\bauthor{\bsnm{Osgood}, \binits{N.D.}}:
\bctitle{Mathematical modeling of {HPA} axis using particle filter algorithm}.
In: \bbtitle{2018 IEEE International Conference on Healthcare Informatics (ICHI)},
pp. \bfpage{400}--\blpage{402}
(\byear{2018}).
\doiurl{10.1109/ICHI.2018.00073}
\end{bchapter}
\endbibitem

\bibitem[\protect\citeauthoryear{Olff et~al.}{2006}]{HPA-HPT2}
\begin{barticle}
\bauthor{\bsnm{Olff}, \binits{M.}},
\bauthor{\bsnm{Güzelcan}, \binits{Y.}},
\bauthor{\bsnm{{de Vries}}, \binits{G.-J.}},
\bauthor{\bsnm{Assies}, \binits{J.}},
\bauthor{\bsnm{Gersons}, \binits{B.P.R.}}:
\batitle{{HPA}- and {HPT}-axis alterations in chronic posttraumatic stress disorder}.
\bjtitle{Psychoneuroendocrinology}
\bvolume{31}(\bissue{10}),
\bfpage{1220}--\blpage{1230}
(\byear{2006})
\doiurl{10.1016/j.psyneuen.2006.09.003}
\end{barticle}
\endbibitem

\bibitem[\protect\citeauthoryear{Rousseau et~al.}{2015}]{Rousseau2015}
\begin{barticle}
\bauthor{\bsnm{Rousseau}, \binits{E.}},
\bauthor{\bsnm{Joubert}, \binits{M.}},
\bauthor{\bsnm{Trzepla}, \binits{G.}},
\bauthor{\bsnm{Parienti}, \binits{J.-J.}},
\bauthor{\bsnm{Freret}, \binits{T.}},
\bauthor{\bsnm{Vanthygem}, \binits{M.-C.}},
\bauthor{\bsnm{Desailloud}, \binits{R.}},
\bauthor{\bsnm{Lefebvre}, \binits{H.}},
\bauthor{\bsnm{Coquerel}, \binits{A.}},
\bauthor{\bsnm{Reznik}, \binits{Y.}}:
\batitle{Usefulness of time-point serum cortisol and {ACTH} measurements for the adjustment of glucocorticoid replacement in adrenal insufficiency}.
\bjtitle{PLoS ONE}
\bvolume{10}(\bissue{8}),
\bfpage{0135975}
(\byear{2015})
\doiurl{10.1371/journal.pone.0135975} .
\bcomment{PHAD Study Group}
\end{barticle}
\endbibitem

\bibitem[\protect\citeauthoryear{Sriram et~al.}{2012}]{Sriram}
\begin{barticle}
\bauthor{\bsnm{Sriram}, \binits{K.}},
\bauthor{\bsnm{Rodriguez-Fernandez}, \binits{M.}},
\bauthor{\bsnm{III}, \binits{F.J.D.}}:
\batitle{Modeling cortisol dynamics in the neuro-endocrine axis distinguishes normal, depression, and post-traumatic stress disorder {(PTSD)} in humans}.
\bjtitle{PLoS Computational Biology}
\bvolume{8}(\bissue{2}),
\bfpage{1002379}
(\byear{2012})
\doiurl{10.1371/journal.pcbi.1002379}
\end{barticle}
\endbibitem

\bibitem[\protect\citeauthoryear{Speer et~al.}{2019}]{SPEER2019100180}
\begin{barticle}
\bauthor{\bsnm{Speer}, \binits{K.E.}},
\bauthor{\bsnm{Semple}, \binits{S.}},
\bauthor{\bsnm{Naumovski}, \binits{N.}},
\bauthor{\bsnm{D'Cunha}, \binits{N.M.}},
\bauthor{\bsnm{McKune}, \binits{A.J.}}:
\batitle{{HPA} axis function and diurnal cortisol in post-traumatic stress disorder: A systematic review}.
\bjtitle{Neurobiology of Stress}
\bvolume{11},
\bfpage{100180}
(\byear{2019})
\doiurl{10.1016/j.ynstr.2019.100180}
\end{barticle}
\endbibitem

\bibitem[\protect\citeauthoryear{Xu et~al.}{2017}]{HPA-HPT}
\begin{barticle}
\bauthor{\bsnm{Xu}, \binits{Y.}},
\bauthor{\bsnm{Liang}, \binits{J.}},
\bauthor{\bsnm{Cao}, \binits{Y.}},
\bauthor{\bsnm{Shan}, \binits{F.}},
\bauthor{\bsnm{Liu}, \binits{Y.}},
\bauthor{\bsnm{Xia}, \binits{Q.}}:
\batitle{High levels of nesfatin-1 in relation to the dysfunction of the hypothalamic-pituitary-adrenal and hypothalamus-pituitary-thyroid axes in depressed patients with subclinical hypothyroidism}.
\bjtitle{Neuropsychiatric Disease and Treatment}
\bvolume{13},
\bfpage{1647}--\blpage{1653}
(\byear{2017})
\doiurl{10.2147/NDT.S138954}
\end{barticle}
\endbibitem

\end{thebibliography}
\end{document}